\documentclass[reprint,amsmath,amssymb,aps,11pt]{revtex4}
\usepackage{graphicx}
\usepackage{bm}
\usepackage{color}
\usepackage{amssymb}
\usepackage{amsmath}
\usepackage{cancel}
\usepackage{ulem}

\begin{document}








\title{Dynamics of $2$D fluid in bounded domain via conformal variables}

\author{Alexander Chernyavsky}
\author{Sergey Dyachenko}
\affiliation{
Department of Mathematics, SUNY at Buffalo
}

\date{\today}

\begin{abstract} 

In the present work we compute numerical solutions of an integro-differential equation for traveling waves 
on the boundary of a $2$D blob of an ideal fluid in the presence of surface tension. We find that 
 solutions with multiple lobes tend to approach Crapper capillary waves in the limit of many lobes. Solutions 
with a few lobes become elongated as they become more nonlinear. It is unclear whether there is a 
limiting solution for small number of lobes, and what are its properties.
Solutions are found from solving a nonlinear pseudo--differential equation by means of the Newton-Conjugate
Residual method. We use Fourier basis to approximate the solution with the number of Fourier modes up to 
$N = 65536$. 
\end{abstract}

\maketitle

\section{Introduction}

The capillary waves are commonly observed in the ocean swell on the length scale of a few centimeters. 
They often appear as a result of breaking of steep gravity waves~\cite{longuet1963generation} or
other processes related to water wave turbulence, such as the formation of direct energy 
cascade~\cite{dyachenko2004weak,korotkevich2008simultaneous}. When wind blows on the ocean surface, or a 
breaking ocean wave is observed, a spray of droplets often is formed when tiny droplets detach from the main body
of water. 
%

Leaving aside the ballistic motion induced by gravity, the internal motion of a droplet is dominated by 
the kinetic energy entrapped in its body, and the surface tension forces acting upon its free surface.
The perfect sphere is a minimizer of the surface area of a fluid droplet and hence potential energy due to 
surface tension attains its minimum value in a spherical droplet. Similarly, in a $2$D fluid the minimal 
perimeter is attained by a disc-shaped droplet. Once detached from the bulk of the fluid, it carries 
away angular momentum from the fluid body as a part of its kinetic energy. Both quantities are conserved 
in time and contribute to its dynamics which can be quite complicated, and is described by the Euler 
equations with a free boundary. 
%
When the angular momentum carried by a droplet is too large to be balanced by surface tension, the droplet may break 
further into even smaller droplets, and a cascade of multiple breaking droplets observed on a large scale may be 
contributing to generation of water spray at the crests of steep ocean waves. Some environmental processes, 
like the gas exchange at the air--water interface, can be greatly enhanced by an effective increase of surface 
area through droplet generation. It is our aim to understand the internal dynamics of droplets, and the processes 
that may result in their breaking.

In the present work, we consider nonlinear traveling waves on the free surface of a droplet of radius $R$ 
induced by a balance of angular momentum and the forces of surface tension.
We follow the conformal variables approach originally introduced by Stokes in~\cite{Stokes1880} and
later extended to time-dependent problem in~\cite{Ovsyannikov1973,Tanveer1991}. 
A framework for studying flows in closed domains (like a fluid blob) as well as on the 
exterior of an air bubble submerged in a 2D ideal fluid has been developed in~\cite{Crowdy1999}.
Our approach is based on the conformal variables technique for bounded domains described in~\cite{dyachenko2022traveling}.

We find families of nonlinear traveling waves that bifurcate from a disc shaped droplet and can be parameterized by 
an integer number of wavelengths per perimeter of a droplet (the number of lobes, $k$), and the wave steepness.
The solutions are found numerically in terms of Fourier series satisfying pseudo--differential nonlinear eigenvalue problem 
that is qualitatively similar to the Babenko equation for Stokes waves in $2$D~\cite{babenko1987some} and can be solved by 
similar numerical techniques~\cite{yang2009newton,Saad2003}. 

Given wave length $\lambda$, we recover the family of Crapper wave solutions~\cite{Crapper1957,Longuet1988} in the 
limit $\lambda\ll R$ ($k\to\infty$). This limit is equivalent to traveling capillary waves in infinite depth fluid. 
Oscillations of 3D droplets have been a subject of interest since the works of Rayleigh~\cite{Rayleigh1879}, and have been 
studied both experimentally and analytically, see e.g.~\cite{Trinh1982}. It is yet unclear whether there is a relation 
between dynamics of $2$D and $3$D droplets.

The text is organized as follows. 
In the first three sections we describe the motion integrals relevant to 
droplet mechanics, the conformal variables approach and the equation describing a nonlinear traveling wave~\eqref{traveling_cmplx}.
Section~\ref{section:series} describes the series solution for infinitesimal waves which are employed as the 
initial guess in the Newton-Conjugate-Residual method~\cite{yang2009newton,Saad2003} in section~\ref{section:nonlinear_waves}.
The main results and conclusions are described in section~\ref{section:conclusion}.

\section{Problem Formulation}

We consider the motion of a $2$D ideal fluid in a bounded domain $\mathcal{D}$. 
The velocity field is given by the gradient of the velocity potential, $\varphi({\bf r}, t)$,
where ${\bf r} = (x,y)^T\in\mathcal{D}$.

The Hamiltonian is the sum of kinetic and potential energy due to the surface tension:
\begin{align}
\mathcal{H} = \frac{1}{2}\iint\limits_{D} \left(\nabla \varphi \right)^2\,dxdy + \sigma 
\int\limits_{\partial D} dl,
\end{align}
where $\nabla$ is the $2$D gradient, and $\sigma$ is the surface tension coefficient.
The boundary of the fluid domain, $\partial\mathcal{D}$, also known as the free surface, 
 is a time-dependent curve in $2$D. 

When the fluid is at rest the shape of the droplet is a perfect disc, a shape that 
attains the least perimeter given a fixed volume $\mu$; when detached from 
the body of fluid, droplet carries away angular momentum $\mathcal{J}$ which 
is conserved:
\begin{align}
\mu = \iint\limits_{D} dxdy \quad\mbox{and}\quad 
\mathcal{J} = \iint\limits_{D} \left[{\bf r} \times \nabla \varphi\right] \, dx dy.\label{mass_cons}
\end{align}


A semi--infinite periodic strip  $w=u+iv\in\{ -\pi \leq u < \pi, v \leq 0\}$ 
is conformally mapped to the fluid domain $z=x+iy\in D$ by the complex function $z(w,t)$.
The invariant quantities associated with the flow may be expressed in terms of 
conformal variables as 1$D$ integrals over 
the free surface $w=u$ ($v=0$); for example, the Hamiltonian becomes
\begin{align}
\mathcal{H} = \frac{1}{2}\iint\limits_D \left( \nabla \varphi \right)^2 \,dxdy + 
\sigma \int\limits_{\partial D}\,dl = \frac{1}{2}\int\limits_{-\pi}^{\pi} \psi\hat k \psi\,du
 + \sigma \int\limits_{-\pi}^{\pi} |z_u| \, du, \label{hamiltonian_gen}
\end{align}
where $\hat H$ is the Hilbert transform, and $\hat k = -\hat H\partial_u$. 
Here $\psi(u,t) = \varphi(x(u,t), y(u,t), t)$ 
is the velocity potential at the free surface. 
The total volume of an incompressible fluid is proportional to the mass of the fluid $\mu$, a trivial 
constant of motion, and the angular momentum $\mathcal{J}$, given by
\begin{align}
&\mu = \iint\limits_{D} \,dx\,dy = \frac{1}{4i}\int\limits_{-\pi}^{\pi} \left[ z \bar z_u - \bar z  z_u\right] \, du, \label{fluid_mass} \\
&\mathcal{J} = \iint\limits_{D} \left[{\bf r} \times \nabla \phi \right] \,dxdy =
-\frac{1}{2}\int\limits_{\partial D} r^2 \frac{\partial \theta}{\partial {\bf n}} \, dl =  -\frac{1}{2}\int\limits_{-\pi}^{\pi} |z|^2 \psi_u \,du, \quad \frac{d\mathcal J}{dt} = 0, \label{ang_mom}
\end{align}
where $r = \sqrt{x^2 + y^2}$ and ${\bf n}$ is the unit normal to 
the free surface.


\section{Traveling Wave}
The implicit form of complex equations of motion is given by:
\begin{align}
&z_t \bar z_u - \bar z_t z_u = \bar \Phi_u - \Phi_u,\label{complex_kinematic} \\
&\psi_t\bar z_u - \psi_u \bar z_t + \frac{\Phi_u^2}{2z_u} = i\sigma \partial_u\left(\frac{\bar z_u}{|z_u|} \right),\label{complex_dynamic}
\end{align}
where $\Phi = \psi + i\hat H \psi$ is the complex potential.



A traveling wave on the free surface of a disc is obtained by seeking conformal map and potential in the form:
\begin{align}
z(u,t) = e^{-i\Omega t} z\left(u - \Omega t \right) \quad\mbox{and}\quad
\Phi(u,t) = i\Omega \hat P|z|^2 - \beta t, \label{z_travel}
\end{align}
where $\beta$ is the Bernoulli constant. We note that the equations of 
motion are invariant under the change of variables $u \to u - \Omega t$, and 
thus the solution may be sought in the form $z = z(u)$. Substitution of~\eqref{z_travel} in the 
equations~\eqref{complex_kinematic} and~\eqref{complex_dynamic} leads to 
an equation for traveling waves:
\begin{align}
2 \beta y_u - \frac{\Omega^2}{2}\left[ x\hat k|z|^2 - \hat H\left(y\hat k |z|^2 \right)  \right] 
- \sigma \partial_u \left[\frac{x_u}{|z_u|} - \hat H \left( \frac{y_u}{|z_u|} \right) \right] = 0,\label{traveling_real}
\end{align}
or, in the complex form,
\begin{align}
2i \beta z_u + \Omega^2 \hat P\left[ z \hat k |z|^2 \right] + 2\sigma \partial_u \hat P\left[ \frac{z_u}{|z_u|} \right] = 0. \label{traveling_cmplx} 
\end{align}
For a traveling wave solution, kinematic constants are related via the formula 
\begin{align}
\mu\beta = \Omega \mathcal{J} + \frac{\sigma}{2} \langle |z_u| \rangle,
\end{align}
where angular brackets denote integral over one period, and the last term is the perimeter of 
the droplet.

\section{\label{section:series}Asymptotic series for small waves}

Let $w = u+iv\in \mathbb{C}^-$, and recall that $e^{-iw}$ is a conformal map from a semi-infinite 
strip $-\pi < u < \pi$ and $v < 0$ to a unit disc. The function $z(u)$ describing the shape of a small 
amplitude wave is represented by an infinite Fourier series,
\begin{equation}
z(u) = e^{-iu}\left( 1 + \sum\limits_{k=1}^{\infty} a_k e^{-iku}\right),\label{series_z}
\end{equation}
where $a_k$ are the Fourier coefficients. Unless the solution is strongly nonlinear the 
series is rapidly convergent, and asymptotic solution of the equation~\eqref{traveling_cmplx}
can be obtained by a series expansion~\eqref{series_z} assuming $|a_2| \ll |a_1|$.  
%
%

The first-order approximation is given by
\begin{equation}
z = e^{-iu}\left(1 + a_k e^{-iku}\right), \label{zfo}
\end{equation}
where $k\geq2$ is an integer representing the number of lobes in the solution. 
When ansatz~\eqref{zfo} is plugged into the dynamic condition~\eqref{complex_dynamic}, 
we find that the Bernoulli constant, $\beta$, and frequency, $\Omega$, must be expanded in 
series keeping $O(a_k)$ terms as follows,
\begin{equation}
\beta = \sigma + O(a_k^2), \quad \Omega^2 = \frac{(k^2-1)\sigma}{k} + O(a_k^2).
\label{params_lin}
\end{equation}
The dispersion relation of linear waves $\Omega(k)$ was also obtained in Ref.~\cite{dyachenko2022traveling}.


The second order approximation to the solutions of~\eqref{traveling_cmplx} 
can be found by keeping the first two terms in the Fourier expansion~\eqref{series_z}.
We will consider the case of $k=2$, but generalization to arbitrary number 
of lobes $k$ can be done analogously. We now seek solution in the form:
\begin{equation}
z=e^{-iu}\left(1 + a_2 e^{-2iu} + a_4 e^{-4iu} \right),
\label{Stokes-series}
\end{equation}
and substitute it into the equation~\eqref{traveling_cmplx} for traveling waves 
to match the corresponding terms in the expansion. We expand the equation~\eqref{traveling_cmplx}
in Fourier series and require that the first three series' coefficients 
vanish, which results in the following expressions:
\begin{align*}
&e^{-iu} \left[ 2(\beta-\sigma) 
	+ \frac{1}{128}\bigl(64(9\sigma+4\Omega^2)a_2^2\bigr) + O(a_2^4)\right] 
	&&\text{(determines $\beta$)} \\
&+e^{-3iu} a_2\left[(6\beta - 9\sigma + 2\Omega^2) 
	+ \left(\frac{9}{8}(-9a_2^2+20a_4)\sigma + 4a_4\Omega^2\right) + O(a_2^4)\right] 
	&&\text{(determines $\Omega^2$)}\\
&+e^{-5iu} \left[128\bigl(5(8a_4\beta + 9a_2^2\sigma  - 20a_4\sigma) + 8(a_2^2+2a_4)\Omega^2\bigr) 
	+ O(a_2^4)\right], &&\text{(determines $a_4$)}
\end{align*}
thus the second order approximation for the Bernoulli constant $\beta$ and frequency $\Omega$ are
determined by 
\begin{align*}
\beta = \sigma \left( 1 - \frac{15}{4}a_2^2\right) + O(a_2^4), \quad 
\Omega^2 = \sigma \left( \frac{3}{2} - \frac{25}{4}a_2^2\right) + O(a_2^4),   
\end{align*}
and $a_4 = \frac{19}{12}a_2^2$. The first and second order approximations are used 
to provide the initial guess for the Newton's method applied to the fully nonlinear
equation~\eqref{traveling_cmplx}. The discussion of fully nonlinear solutions 
and implementation of Newton's method is presented in the following sections. 


\begin{figure}[htbp]
\begin{center}
\includegraphics[width=.45\textwidth]{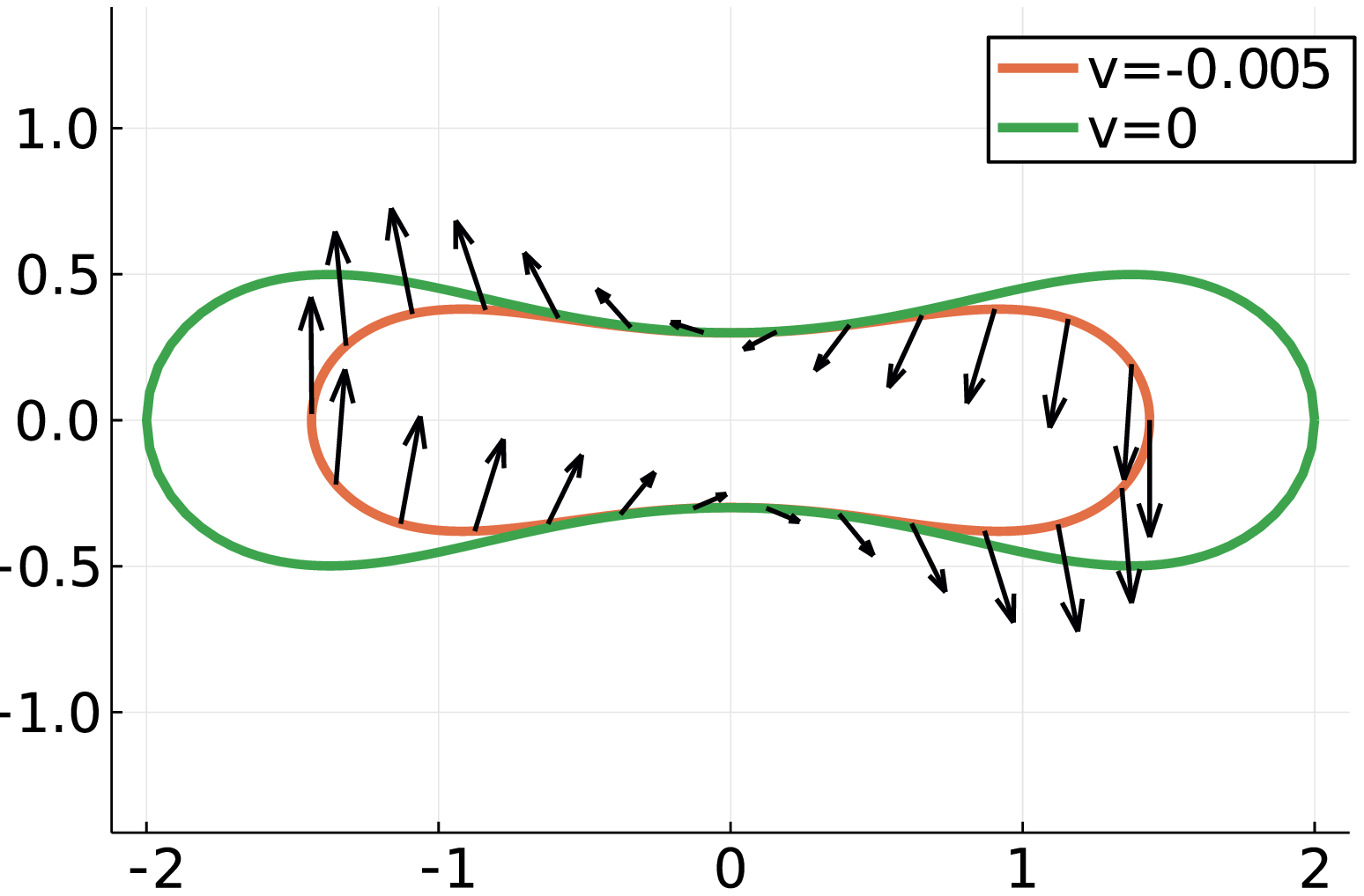}
\includegraphics[width=.45\textwidth]{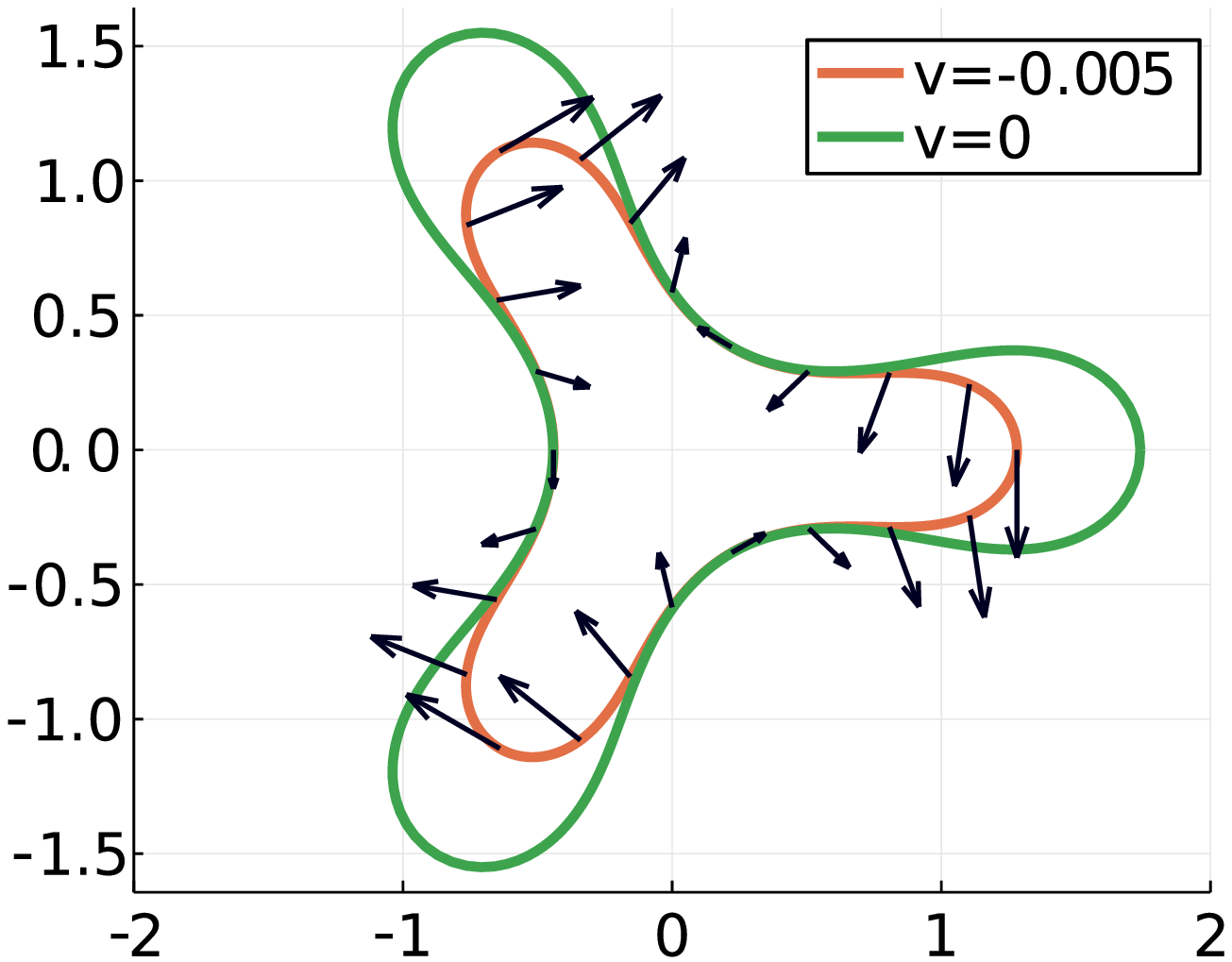}
\caption{
(Left Panel) shows a two-lobed ($k = 2$) solution with $H/\lambda \approx 0.54$, 
and the (Right Panel) shows a three-lobed ($k = 3$) solution with $H/\lambda \approx 0.62$.
The droplet shape is marked by green line, and the orange line corresponds 
to a line inside the fluid at $v = -0.005$. The velocity field is represented by black 
arrows. 
}
\label{series-figure}
\end{center}
\end{figure}

\begin{figure}[htbp]
\begin{center}
\includegraphics[width=.45\textwidth]{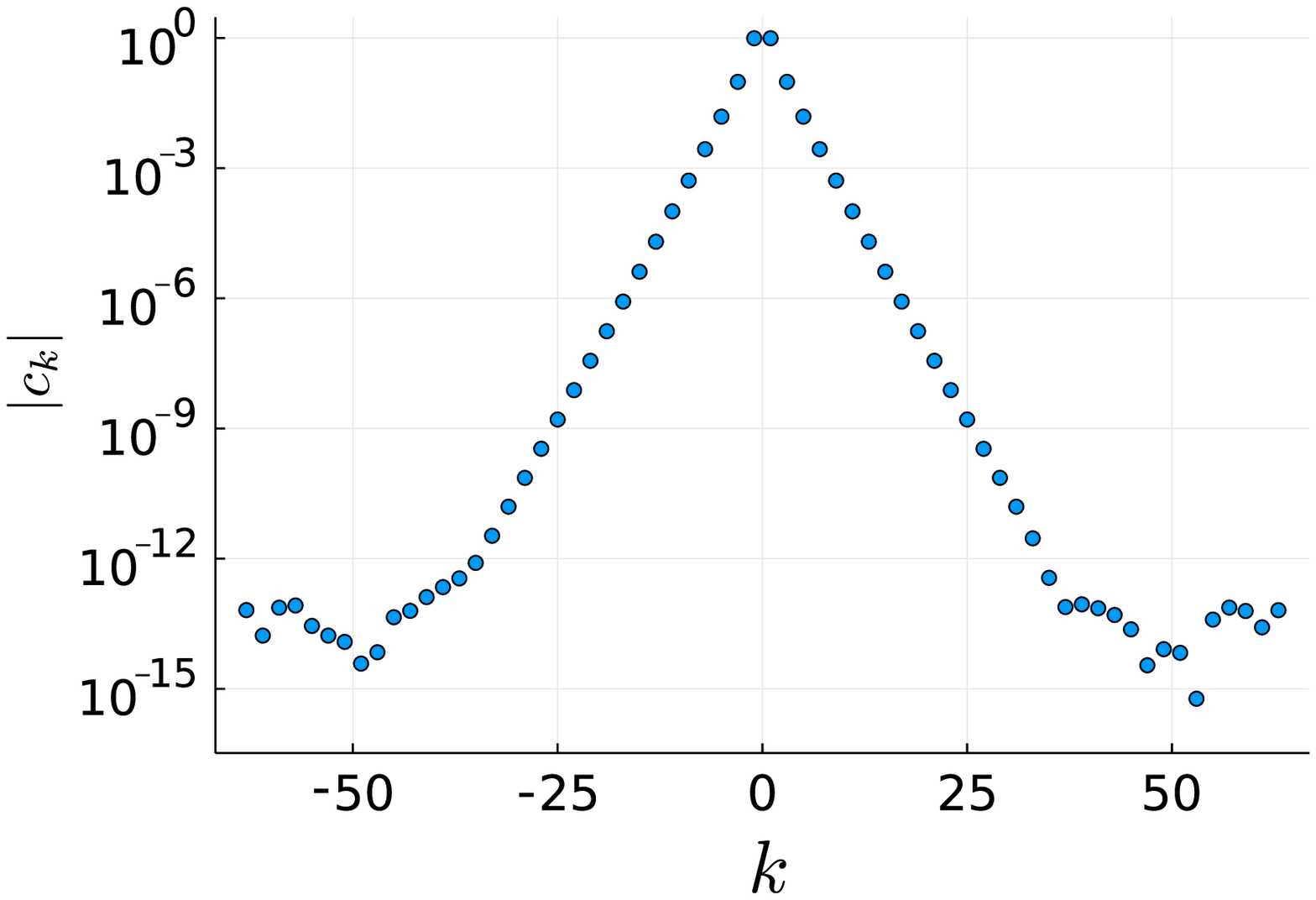}
\includegraphics[width=.45\textwidth]{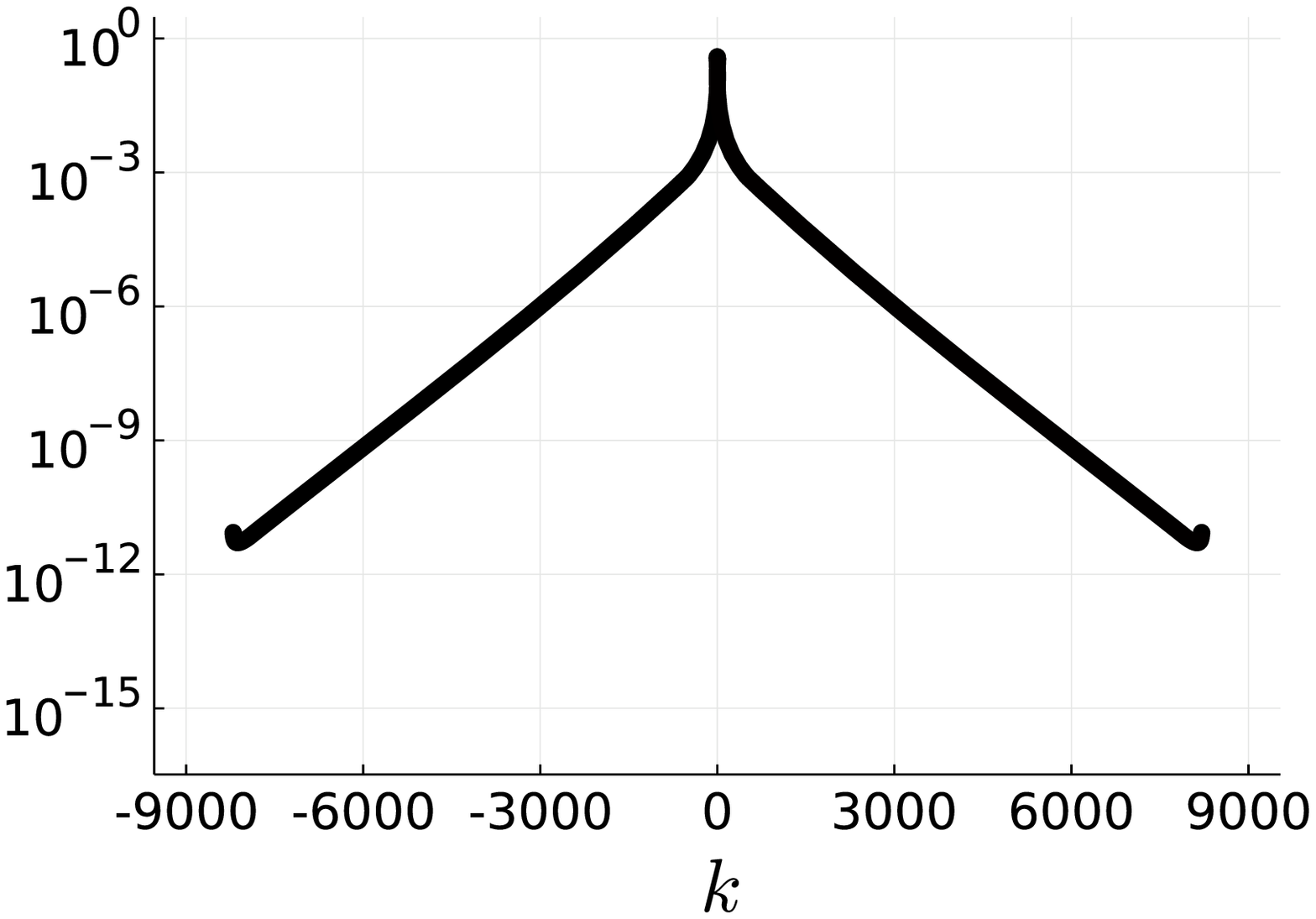}
\caption{
The Fourier spectrum for two nonlinear waves with two lobes ($k = 2$);
(Left Panel) shows the magnitude of Fourier coefficients for a wave 
with $H/\lambda \approx 0.06$, and (Right Panel) corresponds to $H/\lambda \approx 0.54$.
}
\label{spectrum-figure}
\end{center}
\end{figure}

In a more general case, $k\geq2$, the second order approximation 
is found analogously: 
\begin{equation}
z = e^{-iu}(1+a_k e^{-iku}+ a_{2k} e^{-2iku}) + O(a_k^3), 
\label{z-series}
\end{equation}
and by repeating the same steps to keep orders up to $O(a_k^2)$, we obtain
\begin{subequations}
\label{betaOmegac-series}
\begin{align}
\beta &= \left(1 - \frac{a_k^2}{4}(k+1)(3k-1)\right)\sigma + O(a_k^4), \\
\Omega^2 &= 
\frac{k^2-1}{4k(1+2k^2)}
\left(4+8k^2 + a_k^2 (1+k)(6-k(22+k+k^2))\right)\sigma + O(a_k^4), \\
a_{2k} &= \frac{(1+k)((7+2k)k-3)}{4(1+2k^2)}a_k^2 + O(a_k^4).
\end{align}
\end{subequations}
Conformal map~\eqref{z-series} with parameters~\eqref{betaOmegac-series} 
can serve as an initial guess for numerical iterative procedure such as 
Newton's method. 

\section{\label{section:nonlinear_waves}Newton's method}

The nonlinear solutions of~\eqref{traveling_cmplx} are found by 
applying Newton iterations. Given an initial guess, $z^{(0)}(u)$,
or an approximation, $z^{(m)}$ at an iteration $m$, we may write 
the exact solution of~\eqref{traveling_cmplx} as follows,
\begin{equation}
z = z^{(m)} + \delta z, \label{newt1}
\end{equation}
where $z$ is the unknown exact solution of~\eqref{traveling_cmplx}, $m\geq 0$ is the
iteration number, and $\delta z$ is the correction term to be determined.
The formula~\eqref{newt1} is substituted into the equation~\eqref{traveling_cmplx}
assuming $||\delta z|| \ll ||z^{(m)}||$ and only the linear terms in $\delta z$
are kept:
\begin{align*}
2i\beta \delta z_u + \Omega^2 \hat{P}\left[
	\delta z \hat{k} \left|z^{(m)}\right|^2 + z^{(m)}\hat{k}\left(z^{(m)}\delta \bar{z} + \bar{z}^{(m)}\delta z\right)\right] \qquad\,\quad\\
+ \sigma \partial_u \hat{P} \left[ \frac{z^{(m)}_u}{\left|z^{(m)}_u\right|^3} 
	\left( \bar{z}_u^{(m)}\delta z_u 
		- z^{(m)}_u \delta \bar{z}_u\right)\right] + \hat N(z^{(m)})= 0, \label{cmplx_lin}
\end{align*}
where $N(z)$ is defined as follows:
\begin{align}
\hat N(z) := 2i \beta z_u + \Omega^2 \hat P\left[ z \hat k |z|^2 \right] + 2\sigma \partial_u \hat P\left[ \frac{z_u}{|z_u|} \right], 
\end{align}
which is exactly the left-hand side of the equation~\eqref{traveling_cmplx}. It is often 
more convenient to implement iterations for a real unknown function, and we 
may recall that the components of the conformal map are not independent and 
are related via the Hilbert transform, i.e. since $z=x+iy$ and 
$\delta z = \delta x + i\delta y$, then 
\begin{align*}
x = -\hat H y,\quad \delta x = -\hat H\left[\delta y\right],
\end{align*}
and the linearized equation~\eqref{cmplx_lin} is equivalent to an auxiliary real
equation for one real unknown function $\delta y$ given by
\small
\begin{align*}
&-2\beta \delta y_u 
+ \Omega^2 \left(
-\frac12 (\hat{\mathcal{H}} \delta y) \hat{k} |z|^2 - (\hat{\mathcal{H}} y) 
	\hat{k}((\hat{\mathcal{H}} y)(\hat{\mathcal{H}}\delta y) + y\delta y) 
	- \hat{\mathcal{H}}\left[ \frac12 \delta y \hat{k}|z|^2 
		+ y\hat{k} ((\hat{\mathcal{H}} y)(\hat{\mathcal{H}}\delta y) 
	+ y\delta y)\right]\right) \\
&\qquad+ \sigma\partial_u \left(
 -\frac{y_u}{|z_u|^3}(-(\hat{\mathcal{H}} y)_u\delta y_u + y_u (\hat{\mathcal{H}}\delta y)_u) 
	- \hat{\mathcal{H}}\left[-\frac{(\hat{\mathcal{H}} y)_u}{|z_u|^3}
		((\hat{\mathcal{H}} y)_u\delta y_u - y_u (\hat{\mathcal{H}} \delta y)_u)\right]\right) \\
&\qquad 
-2\beta y_u + \frac12\Omega^2 \left(-(\hat{\mathcal{H}}y)\hat{k}|z|^2 
	- \hat{\mathcal{H}}\left[y\hat{k}|z|^2\right]\right) 
	+ \sigma\partial_u \left(\frac{-(\hat{\mathcal{H}} y)_u}{|z_u|} 
		- \hat{\mathcal{H}}\left[\frac{y_u}{|z_u|}\right]\right) = 0,
\end{align*}
\normalsize
or, writing it in compact form,
\begin{equation}
L_1 (y,\delta y) + L_0(y) = 0. 
\label{L1L0}
\end{equation}
When Hilbert transform is applied to the equation~\eqref{L1L0}, we obtain the linear equation for $\delta y$:  
\begin{equation}
\hat{H}L_1(y,\delta y)\delta y = -\hat{H}L_0(y)
\end{equation}
with a self-adjoint operator with respect to the standard inner product: 
\begin{equation}
(f,g) = \int\limits_{-\pi}^\pi f(x)g(x)dx
\end{equation}
for real-valued $f(x)$ and $g(x)$. A linear system with a self-adjoint operator  
is amenable to be solved by the Conjugate-Residual (CR) method. 
We solve~\eqref{L1L0} numerically using a Fourier pseudospectral method to approximate the function $z$. 
The projection operator $\hat{\mathcal{P}}$, Hilbert transform $\hat{H}$ 
and derivatives with respect to $u$ are applied as Fourier multipliers.
At each Newton iteration 
a new linear system is solved with CR, and Newton iterations are performed until a required 
tolerance, $\varepsilon$, is attained: $\|\hat N\left(z^{(m)}\right)\| \leq \varepsilon$.


The second order approximation~\eqref{Stokes-series} is used as an initial guess, $z^{(0)}$, 
to initiate the Newton's iterations. Once a nonlinear solution is determined with a given 
set of parameters, it is used to follow the solution branch to strongly nonlinear waves by 
parameter continuation, 
either in $\beta$ or $\Omega^2$, while keeping surface tension $\sigma$ fixed.

A constraint relating the physically relevant quantities, 
\begin{equation}
\mu \beta - 4 \Omega J - 2\sigma L = 0,
\end{equation}
is used to determine the value of $\Omega^2$ if $\beta$ is known, or vice versa.
Here $4i\mu=\int_{-\pi}^\pi (z \bar{z}_u - \bar{z} z_u )\,du$ is the droplet area, $L=\int_{-\pi}^\pi |z_u|du$ is the droplet perimeter length, 
and $4J= -\Omega \int_{-\pi}^\pi |z|^2 \hat{k}|z|^2 du$ is the rescaled angular momentum. 

The solutions of~\eqref{traveling_cmplx} enjoy two symmetries, one is related to the 
freedom of choosing the phase shift in the rotation angle, and the second symmetry is
related to rescaling of the droplet surface. The choice of phase shift is fixed by 
seeking only even $y(u)$, and in order to hold the droplet area $\mu$ fixed, we rescale 
$z(u)$, $\beta$, $L$ and $J$ once a solution of~\eqref{traveling_cmplx} is 
obtained, 
\begin{equation}
z(u) \rightarrow \frac{z}{\nu}, \quad 
\mu \rightarrow \frac{\mu}{\nu^2}, \quad 
L \rightarrow \frac{L}{\nu}, \quad 
J \rightarrow \frac{J}{\nu^4}. 
\end{equation}
For example, choosing $\nu = \sqrt{\mu/\pi}$ ensures that $\mu=\pi$ is preserved. 
The number of Fourier modes that we considered is limited by $N=65536$, and the 
magnitude of the Fourier mode at series truncation is $10^{-9}$. 
The relative tolerance for solving the linear system~\eqref{L1L0} in the CR method is $10^{-2}$,
and the nonlinear residual for Newton's iterations $\varepsilon = 10^{-9}$.

\section{\label{section:main_results}Main Results}

\begin{figure}[htbp]
\begin{center}
\includegraphics[width=.485\textwidth]{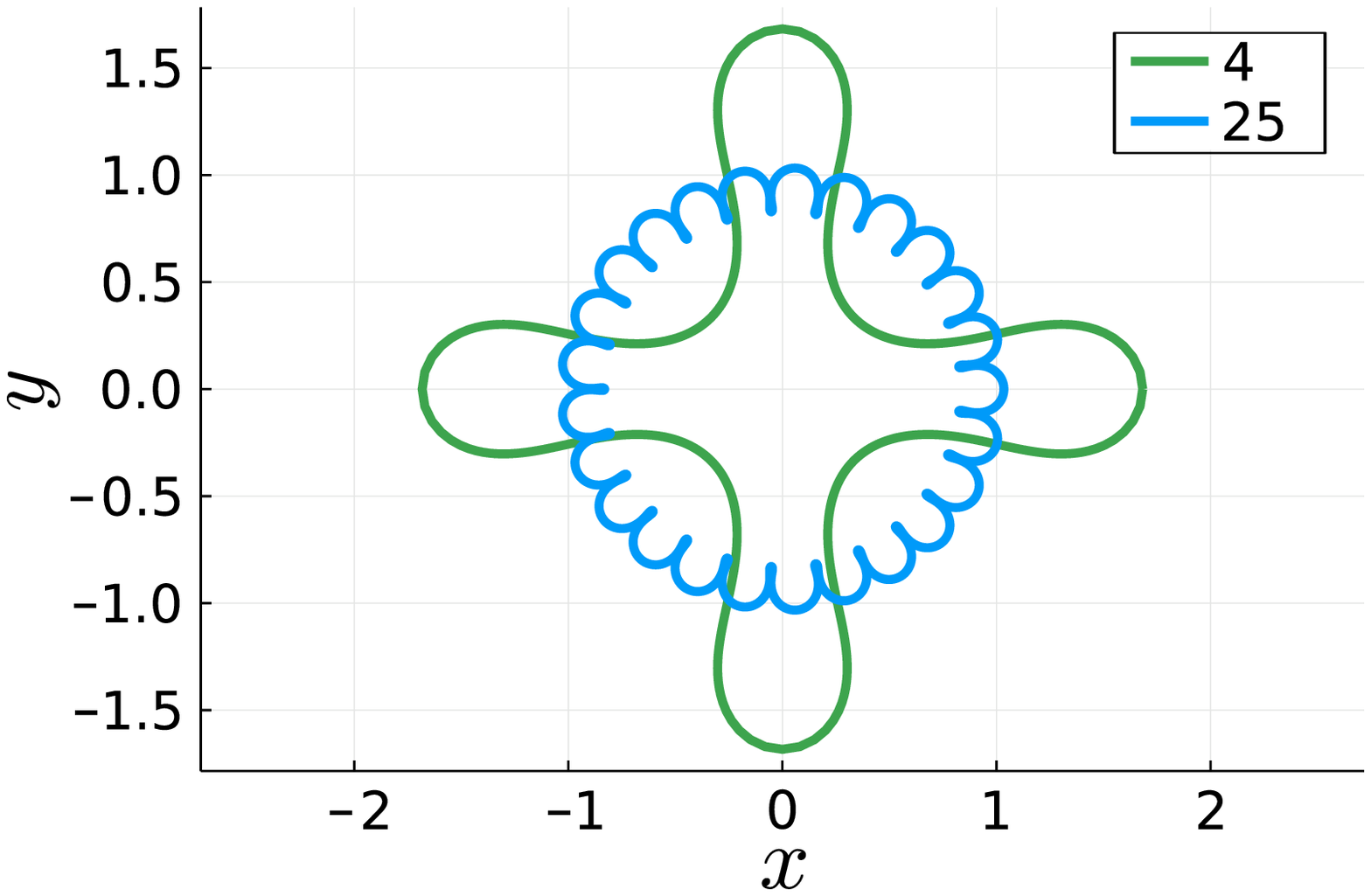}
\includegraphics[width=.485\textwidth]{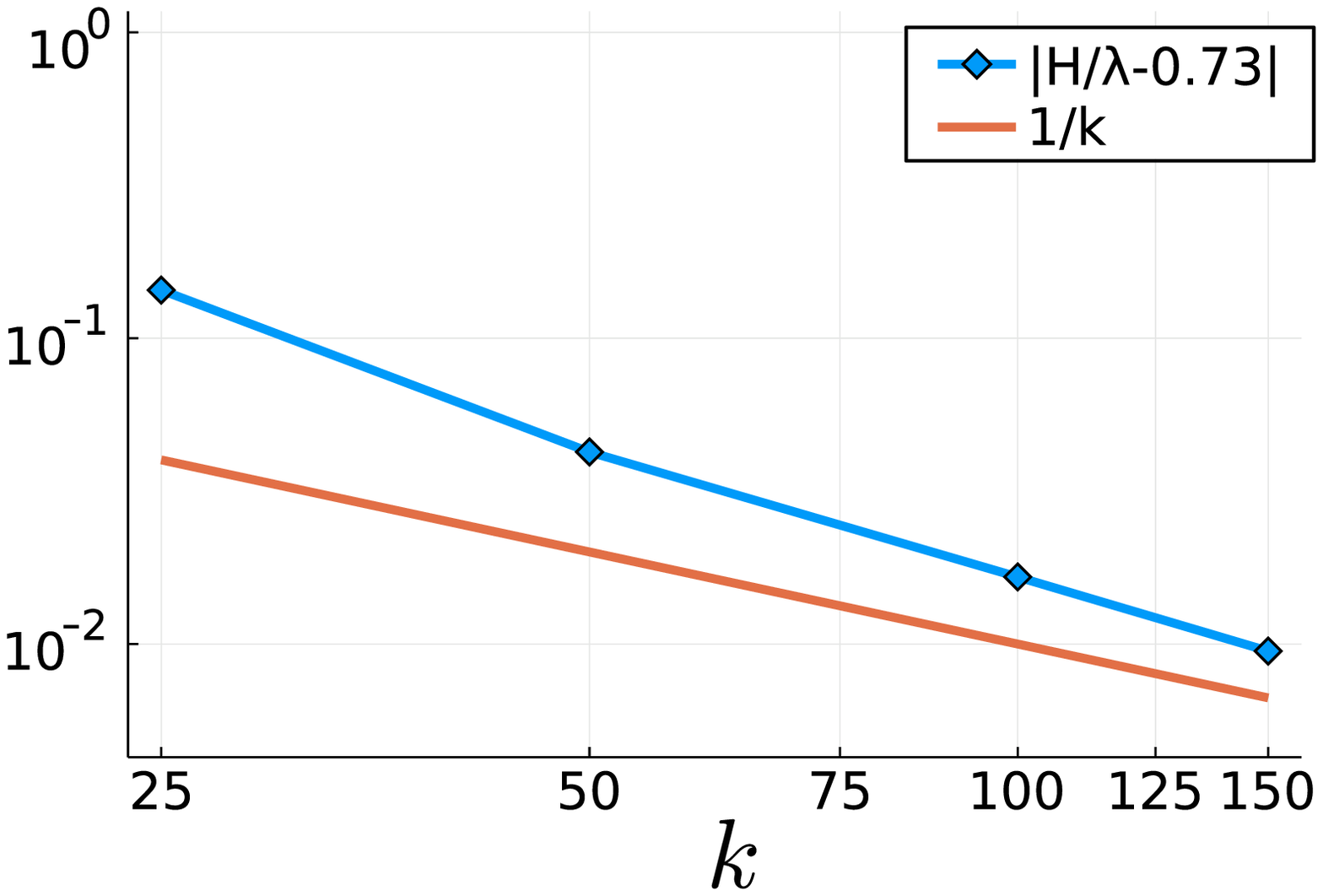}
\caption{
(Left Panel) The shape of a perturbed droplet with $k=4$ and $k=25$. As the number of lobes $k$ grows the 
solution appears to converge to the Crapper wave~\cite{Crapper1957}, and approaches a profile similar to 
the limiting wave with self-intersecting profile. However, when the number of lobes is small the limiting 
scenario remains unclear.
(Right Panel): Wave steepness $H/\lambda$ of self-intersecting solution approaches the value $0.73$ associated 
with the limiting Crapper wave.
}
\label{profiles}
\end{center}
\end{figure}

\begin{figure}[htbp]
\begin{center}
\includegraphics[width=.485\textwidth]{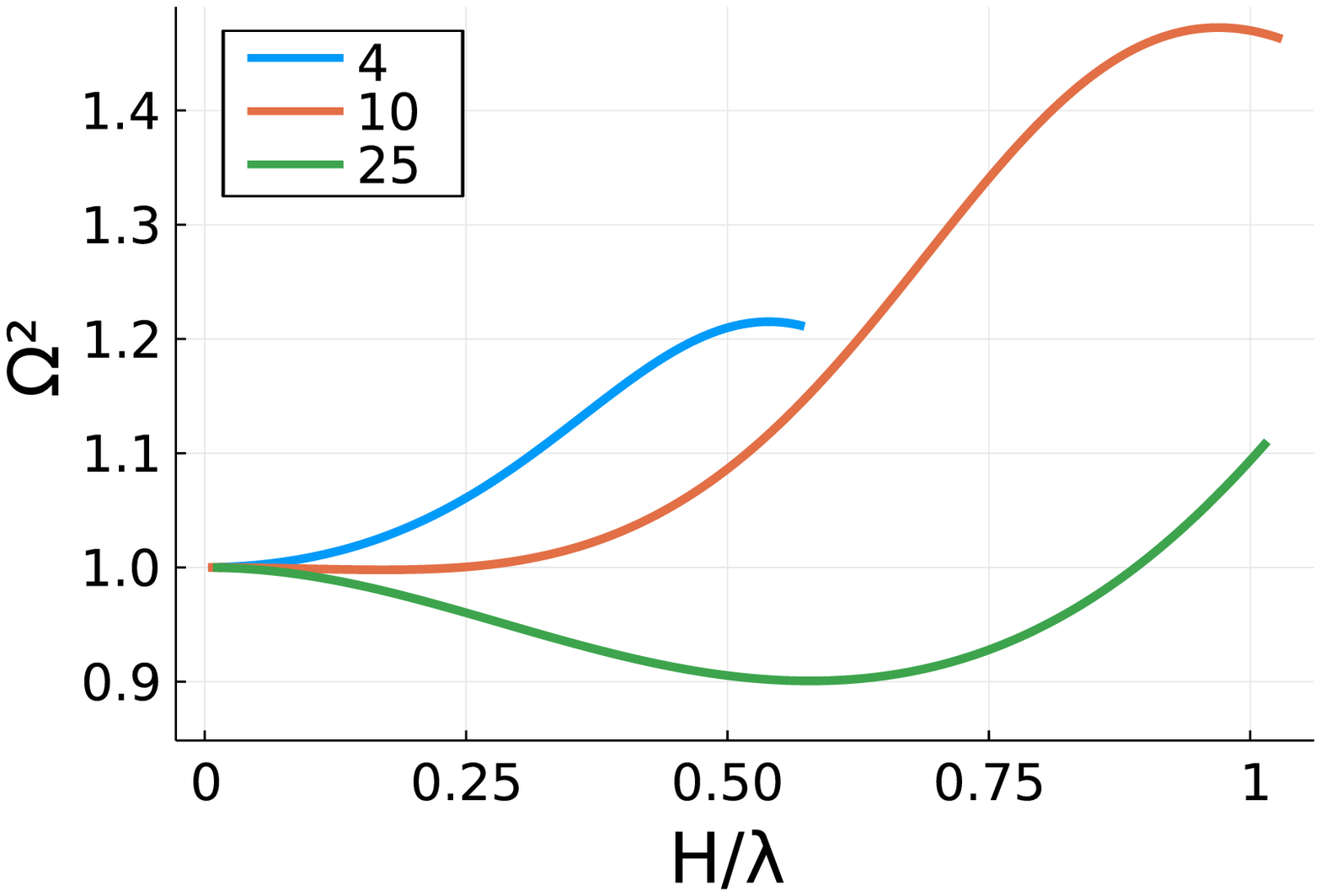}
\includegraphics[width=.485\textwidth]{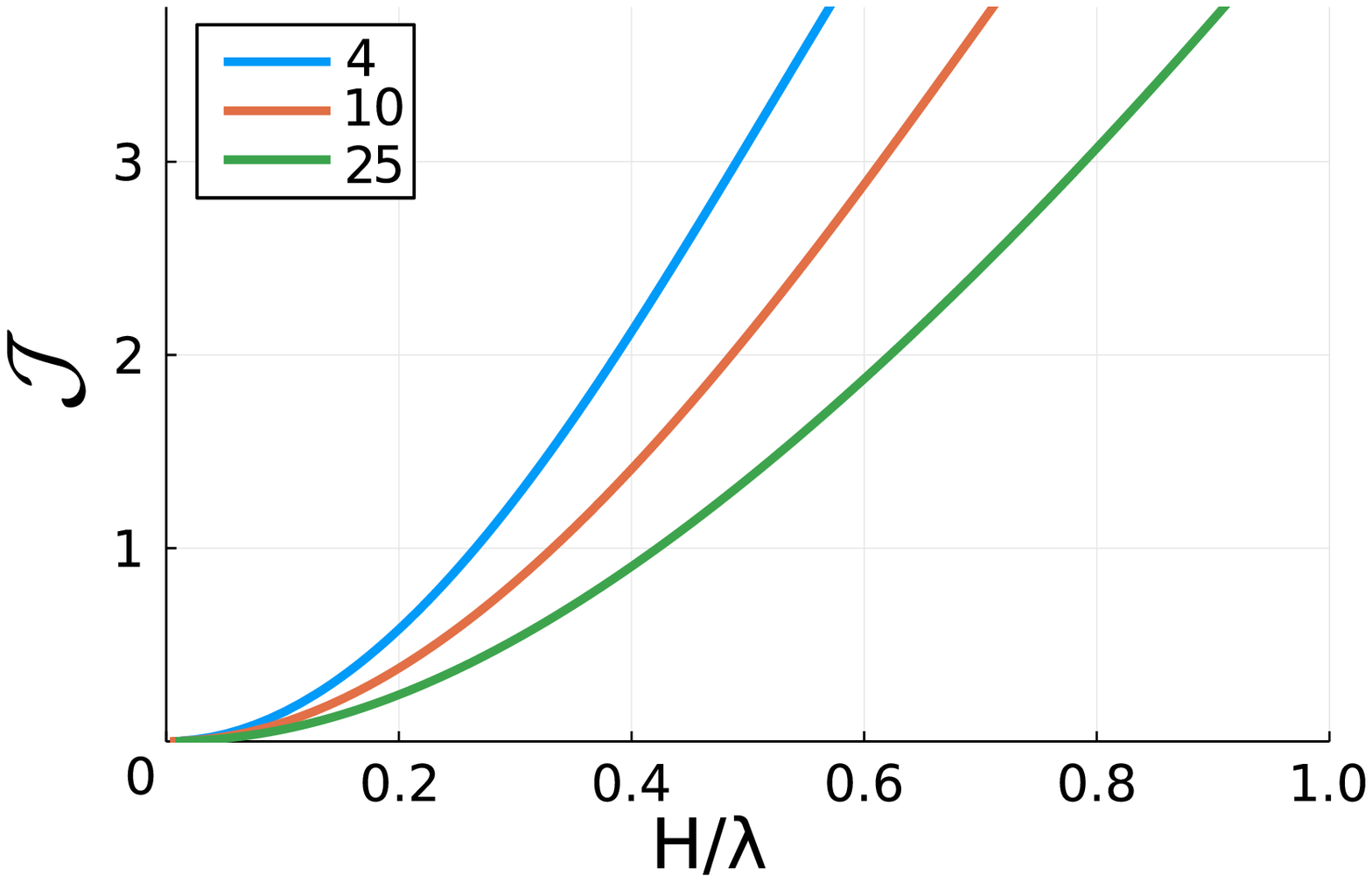}
\caption{
The left panel shows the square of the rotation speed $\Omega^2$ as a function of steepness $H/\lambda$,
and the right panel shows the angular momentum, $\mathcal{J}$, as a function of steepness $H/\lambda$.
}
\label{Hlambda}
\end{center}
\end{figure}

The nonlinear waves obtained with Newton-Conjugate-Residual method are illustrated in 
Fig.~\ref{profiles} with $k=4$ and $k=25$ lobes.
In addition, we show 
parameter curves $\Omega^2$ vs $H/\lambda$ (see Fig.~\ref{Hlambda}, left) 
and $\mathcal{J}$ vs $H/\lambda$ (see Fig.~\ref{Hlambda}, right). Here 
$H$ denotes the height of the wave, and $\lambda$ is the spatial period.

We illustrate two typical solutions of the nonlinear equation~\eqref{traveling_cmplx} 
by showing the shape of the free surface and the velocity field in Fig.~\ref{series-figure}.
We find that a traveling wave becomes elongated as the steepness grows, and the 
number of Fourier modes necessary to resolve the solution grows with wave 
steepness (see Fig.~\ref{spectrum-figure}), indicating the existence of a singularity in the analytic 
continuation of $z(w)$ to the upper half-plane $w \in \mathbb{C}^+$. The nature of this  
singularity, and the existence of a limiting wave is the subject of ongoing research. 

The 
numerical simulations and theoretical considerations suggest that a solution with sufficiently 
many lobes approaches the Crapper wave as the number of lobes grows (Fig.~\ref{profiles}).
The right panel of Fig.~\ref{profiles} indicates that steepness of the self-touching (limiting) Crapper 
wave is approached by the nonlinear solutions of~\eqref{traveling_cmplx} as the number 
of lobes $k$ increases. 
%
%
%
This can be explained as follows: the wavelength is given by $\lambda = 2\pi/k$, and as it becomes small compared to the perimeter of the droplet (when $k$ grows), the effects of local curvature become less significant and vanish in the limit $k\to \infty$.   
%
Another open question concerns the number of lobes for which 
self-touching of neighboring waves occurs (Crapper scenario) versus the presently 
unknown limiting scenario for few lobes, e.g. $k = 2$ and $k = 3$ for which no indication 
of a tendency to self-intersect was observed.  


\section{\label{section:conclusion}Conclusion}
Breaking of water waves in deep ocean is associated with generation of water droplet spray. 
The latter partially accounts for the energy--momentum transfer in wave turbulence. The 
physical processes that generate water spray have been observed in physical 
ocean~\cite{ErininEtAlSprayGeneration2019}, as well as theoretically~\cite{DyachenkoNewell2016}.
As plunging breaker develops on the crest of an ocean wave, there is an abrupt growth of small
scale features, and several physical mechanisms suddenly come into play. The force of the surface 
tension, normally having little effect on long gravity waves becomes one of the dominant 
forces at the crest of a breaking wave. The detachment of a water droplet from a plunging breaker is a 
complicated and nonlinear process, and the present work does not make an attempt to 
understand it to the full extent.

We considered a problem of deformation of a fluid disc with a free boundary subject to the 
force of surface tension. We found that a conformal map associated with such a flow satisfies 
a pseudo--differential equation that is similar to Babenko equation for the Stokes wave. 
We demonstrate the results of numerical simulation with initial 
data close to linear waves, and observe excellent agreement for small amplitude waves, and report 
significant deviations as amplitude grows. 

The nonlinear equation~\eqref{traveling_real}, or its complex form~\eqref{traveling_cmplx},
are solved by the Newton-Conjugate-Residual method~\cite{yang2009newton,Saad2003} that is also applicable to the Stokes wave 
problem. The present work is a precursor to further investigation of nonlinear waves, and 
 of particular interest are the questions 
of existence of the limiting wave, its nature and singularities. One may speculate that 
the limiting wave will not form an angle on the surface, since it would make the potential energy 
grow; yet the numerical simulations suggest the breaking of a droplet (for small number of lobes), and 
a tendency to develop a self-touching solution like the Crapper wave (for large number of lobes). The 
study of limiting scenarios is the subject of ongoing work.   

\newpage



\bibliographystyle{siamplain}
\bibliography{refs.bib}
\end{document}